\documentclass[sigconf]{acmart}

\usepackage{hyperref}
\hypersetup{
  colorlinks   = true, 
  urlcolor     = blue, 
  linkcolor    = blue, 
  citecolor    = blue  
}

\AtBeginDocument{%
  \providecommand\BibTeX{{%
    \normalfont B\kern-0.5em{\scshape i\kern-0.25em b}\kern-0.8em\TeX}}}

\setcopyright{acmcopyright}
\copyrightyear{2024}
\acmYear{2024}
\acmDOI{XXXXXXX.XXXXXXX}

\acmConference[SAC '25]{The 40th ACM/SIGAPP Symposium On Applied Computing
Sicily, Italy, March 31 - April 4, 2025}
\acmBooktitle{The 40th ACM/SIGAPP Symposium On Applied Computing
Sicily, Italy, March 31 - April 4, 2025}
\acmPrice{15.00}
\acmISBN{978-1-4503-XXXX-X/18/06}

\begin{document}

\title{Advancing Quantum Software Engineering: A Vision of Hybrid Full-Stack Iterative Model}


\author{Arif Ali Khan}
\affiliation{%
  \institution{M3S Empirical Software Engineering Research Unit, University of Oulu}
  \city{Oulu}
  \country{Finland}
  \postcode{Oulu FI-90014}}
\email{arif.khan@oulu.fi}

\author{Davide Taibi}
\affiliation{%
  \institution{M3S Empirical Software Engineering Research Unit, University of Oulu}
  \city{Oulu}
  \country{Finland}
  \postcode{Oulu FI-90014}}
\email{davide.taibi@oulu.fi}


\author{Muhammad Azeem Akbar}
\affiliation{%
  \institution{LUT University}
  \city{Lapparante}
  \country{Finland}}
  \postcode{53850 }
  \email{azeem.akbar@lut.fi}


\renewcommand{\shortauthors}{Khan, et al.}

\begin{abstract}
This paper introduces a vision for Quantum Software Development lifecycle, proposing a hybrid full-stack iterative model that integrates quantum and classical computing. Addressing the current challenges in Quantum Computing (QC) such as the need for integrating diverse programming languages and managing the complexities of quantum-classical systems, this model is rooted in the principles of DevOps and continuous software engineering. It presents a comprehensive lifecycle for quantum software development, encompassing quantum-agnostic coding, testing, deployment, cloud computing services, orchestration, translation, execution, and interpretation phases. Each phase is designed to accommodate the unique demands of QC, enabling traditional software developers to engage with QC environments without needing in-depth QC expertise. The paper presents a detailed implementation roadmap, utilizing a range of existing tools and frameworks, thereby making quantum software development more accessible and efficient. The proposed model not only addresses current challenges in quantum software development but also makes a substantial contribution to the field of Quantum Software Engineering (QSE). By proposing a structured and accessible model, it sets the stage for further advancements and research in QSE, enhancing its practicality and relevance in a wide range of applications.
\end{abstract}



\maketitle
\section{Introduction}
Quantum Computing (QC) has recently seen increasing interest from academia and industry. This emerging technology offers groundbreaking opportunities for business growth in fields like cybersecurity, finance, and aerospace. However, the development of cost-effective and reliable quantum software has been progressing slowly. Addressing this challenge is essential for QC to actualize its industrial potential \cite{eskandarpour2020quantum}.
The term "quantum" in Quantum Computing refers to the principles of quantum mechanics, which the system uses for the computation. In physics, "quantum" denotes the smallest amount of any physical entity, typically relating to particles like electrons, neutrons, and photons. QC, a rapidly expanding research area, is anticipated to revolutionize various industries. It leverages the characteristics of quantum mechanics to process information, enabling it to perform certain tasks much faster than classical computers \cite{zhao2020quantum, deutsch1985quantum}. Quantum characteristics such as superposition, entanglement, and quantum interference are utilized in computing. In comparison to classical, quantum computers best perform in several ways. For instance, in areas requiring high-speed parallel computing, such as complex algorithmic problem-solving and advanced computational modeling in fields like astrophysics and climate science. Additionally, they are particularly effective in tasks that demand highly accurate and detailed computations, like simulating molecular interactions in drug discovery and optimizing large-scale logistical operations. Consequently, technology giants such as IBM \cite{qiskt2021}, Google \cite{cirq2022}, and Microsoft \cite{azure2021} are annually investing hundreds of millions of dollars in the development of both quantum hardware and software solutions. These investments are specifically aimed at enhancing the capabilities of QC applications, reflecting the significant potential they offer in various fields.
However, to realize the full potential of QC, effective quantum software solutions are essential, and the development of large-scale quantum software is still far from being a widespread reality. Yet, a novel scientific field, quantum software engineering (QSE) \cite{de2022software}, has emerged to equip developers with the tools to create quantum programs as confidently as they do classical ones. QSE focuses on integrating classical software engineering practices into quantum programming. It particularly focuses on the association of classical and quantum systems, leading to the development of hybrid (quantum/classical) systems. However, various challenges exist in operationalizing the QSE activities based on hybrid practices.
This paper presents a vision of a full-stack quantum software development lifecycle, drawing inspiration from modern-day iterative DevOps capabilities \cite{fitzgerald2017continuous}. However, this approach diverges from traditional DevOps practices due to the intricate nature of integrating classical and high-performance computing resources with quantum computations in an iterative manner \cite{microsoft}. This vision aligns with the strategy of incorporating classical software engineering practices into QSE, thereby underpinning the concept of hybrid computing.

\section{Background and Motivation}
The flaky characteristics of QC necessities the development of new techniques, tools, processes, and methods that are specifically designed for developing a quantum software system. Designing quantum software is particularly challenging due to the inherent characteristics of quantum mechanics like superposition and entanglement. There is a strong demand for new principles, processes, and methodologies in QSE, as these are crucial in the development of quantum software systems \cite{zhao2020quantum}. Following the evolution of conventional software programming, which has advanced from the fixed, hardware-based approaches of the 1950s to today's iterative agile development, QSE is also expected to follow a similar roadmap. It's likely to shift towards a more adaptable, repetitive, and gradual approach to development \cite{piattini2021toward} as highlighted in a national agenda for software engineering research and development by SEI-Carnegie Mellon University \cite{carleton2021architecting}: \textit{“the development of Quantum Computing Software Systems is considered a pivotal future focus area. The objectives of this research area are initially focused on facilitating the development of quantum software for existing quantum computers and then enable increasing abstraction as larger, fully fault-tolerant quantum computing systems become available. A key challenge is to eventually, fully integrate these types of systems into a unified classical and quantum software development life cycle”}. However, this doesn’t come without challenges, and various research studies have been conducted to investigate the challenges of quantum software development processes, tools, and methods.

For instance, Weder et al. \cite{weder2020quantum} have identified that quantum-specific complexities in developing software applications necessitate expertise from multiple disciplines, which leads to the rise of QSE. They emphasize that hybrid quantum applications, which integrate quantum and classical tools and methods, require life cycles that cater to both aspects. The study presented a detailed quantum software development life cycle, exploring various phases, tools, and the integration of distinct software artifacts' life cycles. The article provides an overview of the basics of hybrid quantum applications, describes the proposed life cycle, addresses assumptions and limitations, examines related work, and concludes with thoughts on future developments. De Stefano et al. \cite{de2022software} investigate the transition of quantum computing from scientific research to industrial application. They identify the emergence of QSE, necessitating new engineering methods for large-scale quantum software applications. The study includes mining GitHub repositories using key quantum programming frameworks and conducting surveys with developers to understand adoption challenges. Study findings reveal limited adoption and diverse challenges, highlighting the need for broader software engineering approaches in quantum programming, encompassing both technical and socio-technical aspects. Recently, Akbar et al. \cite{akbar2022classical} proposed a readiness model to aid organizations in transitioning from classical to quantum software engineering. This readiness model describes the core process areas, challenges, and enablers in quantum software engineering. It aims to provide a roadmap for organizations, highlighting best practices and key factors for a successful shift to quantum computing, thereby facilitating a strategic approach to this emerging field. Romero-Alvarez et al. \cite{romero2023enabling} highlight the challenges in quantum software development, such as the need for understanding quantum hardware and low-level programming languages. It advocates for integrating software engineering principles to simplify quantum service development, proposing a continuous deployment strategy for these services. The approach includes an extension of the OpenAPI Specification for quantum algorithms, validated through an API implementation and positively evaluated by developers and students.

Based on insights from related work studies, this vision paper proposes a hybrid (quantum, classical/HPC) full-stack iterative quantum software development lifecycle model. Iterative processes are highly effective in identifying and rectifying bugs and issues at an early stage, where they can be resolved more straightforwardly. Classical computing resources, generally being cost-effective, make such iterative processes financially viable. However, the same doesn't hold true for HPC and QC, where resources are often expensive, and operations consume considerable time and energy. To address this, we propose this idea to allow developers to write their code in a quantum-agnostic way and iteratively integrate it into the system with increasing complexity (from classical to quantum), which is continuously managed in the cloud, involving quantum/classical orchestration, quantum code translation, execution, and results interpretation. The primary contribution of this idea is to facilitate developers in creating quantum software within a hybrid environment without requiring extensive knowledge of QC concepts.

\section{Hybrid Full-Stack Iterative Quantum Software Development Model}

\begin{figure*}[h]
  \centering
  \includegraphics[width=0.65\linewidth]{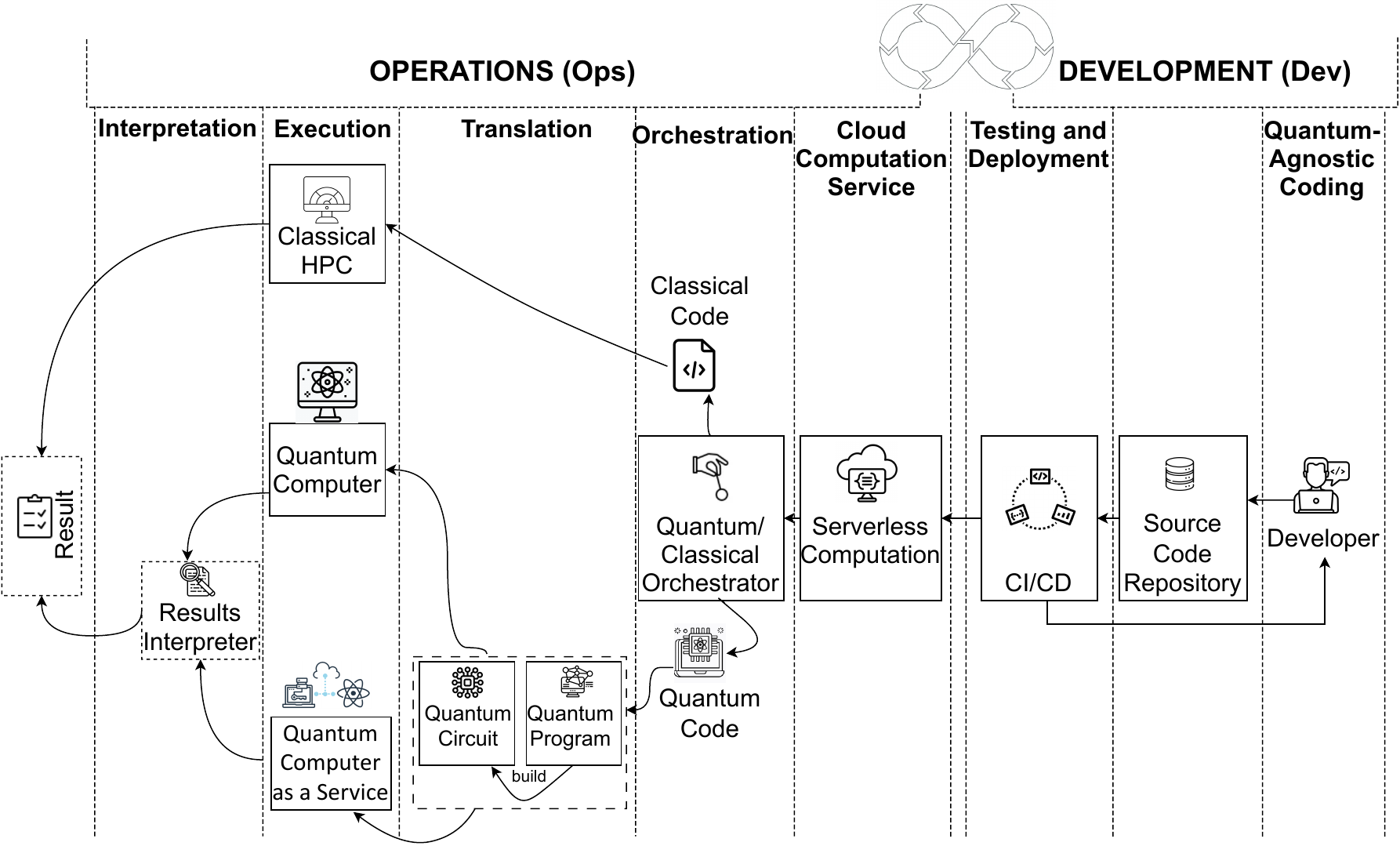}
  \caption{Proposed Model}
  \label{Fig.Proposed-Model}
\end{figure*}

Recent research has revealed that developers working on quantum applications frequently encounter challenges related to the development environment \cite{zhao2020quantum, weder2020quantum, piattini2021toward}. These challenges primarily stem from integrating quantum and classical components. Factors contributing to these issues include different programming paradigms and languages, the complexity of aligning input between classical and quantum segments, and the significant vendor lock-in that characterizes quantum applications \cite{de2022towards}. Based on the principles of iterative DevOps and continuous software engineering \cite{fitzgerald2017continuous}, we aim to address these issues by proposing a hybrid full-stack quantum software development lifecycle model. This model will enable traditional developers to write their own code and integrate it into a quantum environment without worrying about the technical details of QC and the overall infrastructure. Figure \ref{Fig.Proposed-Model} presents the high-level structure of the proposed model, consisting of six core phases: quantum-agnostic coding, testing/deployment, cloud computing services, orchestration, translation, execution, and interpretation. The first two components align with the 'Dev' (Development) aspect of DevOps and the remaining four with 'Ops' (Operations); here is how each phase is sequentially connected and where iterations can be incorporated:
\begin{itemize}
    \item \textbf{Quantum-Agnostic Coding (Dev):} Developers begin by writing code in a quantum-agnostic manner. This is the initial development phase, where the primary focus is on creating the code, which is then stored in a source code repository, serving as a central platform for code management and integration into the CI/CD pipeline. Iteration at this stage involves refining developed code based on feedback and testing outcomes.
    \item \textbf{Testing and Deployment (Dev):} The code moves into the CI/CD pipeline, an inherently iterative process. In CI, the code is regularly merged with the main branch, ensuring compatibility with the existing codebase. This phase includes running automated tests to verify the code functionality, a process that is repeated iteratively to continually enhance code quality. Following this, in CD, each code iteration that successfully passes the tests is automatically deployed to production environments. This repetitive cycle of integration, testing, and deployment underscores the iterative nature of the CI/CD component, facilitating continuous improvement and rapid deployment of changes.
    \item \textbf{Cloud Computing Services (Ops):} This component is critical for integrating and optimizing the deployed code within a cloud environment. This phase provides a scalable, cloud-based platform for deploying the tested code tailored to the needs of developers without deep QC expertise. By leveraging cloud computing, we making quantum software development more approachable for traditional developers. The primary objective is to streamline the quantum software development process, enhancing its efficiency and accessibility, and ensuring smooth transitions between classical and quantum computing paradigms within the development lifecycle.
    \item \textbf{Orchestration (Ops):} It acts as a decision-making junction where the traditional code is directed towards the classical/HPC environment, while the quantum code is channeled into the subsequent translation phase. This orchestration phase essentially functions as a classifier, distinguishing between the code that can be executed using classical/HPC systems and that which needs to be translated into quantum code and circuits for execution on QC systems. The primary purpose of this phase is to efficiently manage and route different segments of code based on their computational requirements. By doing so, it ensures that each type of code is processed in the most suitable environment - classical code in traditional computing infrastructures and quantum code in environments equipped for quantum processing.
    \item \textbf{Translation (Ops):} This component is defined for the quantum-specific code, which was identified and segregated in the previous orchestration phase, undergoes a critical transformation. Here, it is converted into a format that is compatible with quantum hardware. This translation process involves turning the quantum-agnostic code into executable quantum code and circuits, preparing it for implementation on quantum computing systems and simulators. Due to the complex nature of QC, this phase often requires iterative refinement. The translation process must be fine-tuned to ensure both accuracy and compatibility with various quantum systems. This iterative approach is vital to address the unique challenges posed by different quantum hardware architectures and to achieve optimal performance in quantum computations.
    \item \textbf{Execution (Ops):} Distinct paths are followed for the execution of classical and quantum code. Classical code, developed for HPC environments, is executed on robust computing systems. These systems are specifically designed to manage the heavy computational demands typical of classical code segments efficiently. On the other hand, the quantum code, refined and translated in the previous phases, is run on quantum hardware. This hardware may be dedicated quantum computers or accessible quantum computing services offered via cloud platforms. 
    \item \textbf{Interpretation (Ops):} The quantum hardware or simulation output data is processed using various algorithms (e.g... statistical analysis algorithms, classical simulations of quantum code) designed to translate quantum information into classical data formats. These algorithms typically involve statistical analysis and data conversion techniques tailored to make sense of the quantum results in terms of classical computing paradigms. Additionally, the interpreted data is often compared and combined with results from the classical HPC systems to provide a comprehensive overview. This comparative analysis helps in understanding the contributions of both quantum and classical computations to the overall problem-solving process.
    \end{itemize}
    
\section{Execution Roadmap}
To operationalize the proposed model, a comprehensive implementation roadmap utilizing various tools, techniques, and frameworks is essential. This plan will effectively merge classical and quantum computing paradigms into a practical application. Starting with \textit{Quantum-Agnostic Coding} in the Development phase, developers can use quantum programming frameworks like Qiskit \cite{qiskt2021}, Cirq \cite{cirq2022}, or Microsoft Quantum Development Kit \cite{azure2021}. These tools enable writing code adaptable for quantum environments while maintaining familiarity for those skilled in classical programming. Integrating this with version control systems such as GitLab\footnote{https://about.gitlab.com/}  is crucial for effective source code management. The \textit{Testing and Deployment} phase involves setting up CI/CD pipelines using Jenkins\footnote{https://www.jenkins.io/} . This setup facilitates regular code merging and automated testing, ensuring code quality and compatibility. The use of unit testing frameworks like PyTest\footnote{https://docs.pytest.org/en/7.4.x/}  will automate the testing of quantum-agnostic code. In the \textit{Cloud Computing Services } phase, cloud platforms like IBM Cloud could be used for deploying and optimizing the code. The \textit{Orchestration phase}, acting as a decision-making junction, utilizes tools like Kubernetes\footnote{https://kubernetes.io/}. These orchestration tools manage and route code in cloud environments efficiently. Middleware or abstract layer solutions are essential for effective communication between classical and quantum computing environments. In the \textit{Translation} phase, quantum simulators like IBM Qiskit Aer \cite{qiskt2021} can be used for testing the translated quantum code. Tools for converting quantum-agnostic code into executable quantum code and circuits are developed or utilized, ensuring accuracy and compatibility with various quantum systems. The \textit{Execution} phase follows distinct paths for classical and quantum code. Classical code is executed on HPC systems, while quantum code runs on quantum hardware, accessible through services like Rigetti\footnote{https://www.rigetti.com/} Quantum Cloud Services . Finally, the \textit{Interpretation} phase involves processing the quantum hardware or simulation output data. Algorithms for statistical analysis and classical simulations of quantum code are used to translate quantum information into classical data formats. Tools like Python SciPy\footnote{https://scipy.org/}, along with visualization tools like Tableau\footnote{https://www.tableau.com/}, are employed for presenting the integrated results. By adopting Agile methodologies and ensuring cross-disciplinary training, this roadmap offers a structured approach to transform the vision of Hybrid Full-Stack Iterative Quantum Software Development into an operational reality.

\section{Conclusions}
In conclusion, this paper presents a vision of a Hybrid Full-Stack Iterative Quantum Software Development lifecycle, bridging the gap between quantum and classical computing paradigms. Grounded in the principles of DevOps and continuous software engineering, this model addresses the unique challenges of quantum computing, such as integrating diverse programming paradigms and managing the complexities of quantum-classical interaction. By outlining a comprehensive lifecycle that encompasses quantum-agnostic coding, orchestrated deployment, and specialized translation to execution and interpretation phases, the model paves the way for developers to engage with quantum computing more effectively. The proposed roadmap for implementation, leveraging a suite of modern tools and frameworks, aims to make quantum software development accessible and efficient, even for those without extensive quantum computing expertise. This approach not only facilitates the practical realization of quantum software engineering but also sets a foundation for future advancements in this rapidly evolving field.

\bibliographystyle{acm}
\bibliography{references}

\end{document}